\documentclass[%
reprint,
superscriptaddress,
amsmath,amssymb,
prb,
]{revtex4-2}

\usepackage{graphicx}
\graphicspath{{./Images/}}

\usepackage{siunitx}

\usepackage{upgreek}

\usepackage{amsmath}

\usepackage{tcolorbox}
\begin{document}

\title{Local signatures of electron-electron scattering in an electronic cavity}
\author{Carolin Gold}
\email[]{cgold@phys.ethz.ch}
\affiliation{Solid State Laboratory, ETH Zurich, 8093 Zurich, Switzerland}

\author{Beat A. Br\"am}
\affiliation{Solid State Laboratory, ETH Zurich, 8093 Zurich, Switzerland}
\author{Richard Steinacher}
\affiliation{Solid State Laboratory, ETH Zurich, 8093 Zurich, Switzerland}

\author{Tobias Kr\"ahenmann}
\affiliation{Solid State Laboratory, ETH Zurich, 8093 Zurich, Switzerland}
\author{Andrea Hofmann}
\affiliation{Solid State Laboratory, ETH Zurich, 8093 Zurich, Switzerland}

\author{Christian Reichl}
\affiliation{Solid State Laboratory, ETH Zurich, 8093 Zurich, Switzerland}

\author{Werner Wegscheider}
\affiliation{Solid State Laboratory, ETH Zurich, 8093 Zurich, Switzerland}

\author{Mansour Shayegan}
\affiliation{Department of Electrical Engineering, Princeton University, Princeton, New Jersey 08544, USA}
\author{Klaus Ensslin}
\affiliation{Solid State Laboratory, ETH Zurich, 8093 Zurich, Switzerland}
\author{Thomas Ihn}
\affiliation{Solid State Laboratory, ETH Zurich, 8093 Zurich, Switzerland}

\date{\today}

\begin{abstract}
We image equilibrium and non-equilibrium transport through a two-dimensional electronic cavity using scanning gate microscopy (SGM).
Injecting electrons into the cavity through a quantum point contact close to equilibrium, we raster-scan a weakly invasive tip above the cavity regions and measure the modulated conductance through the cavity.
Varying the electron injection energy between $\pm\SI{2}{meV}$, we observe that conductance minima turn into maxima beyond the energy threshold of $\pm\SI{0.6}{meV}$. This observation bears similarity to previous measurements by Jura {\em et al.} [Jura et al., Phys. Rev. B {\bf 82}, 155328 (2010)] who used a strongly invasive tip potential to study electron injection into an open two-dimensional electron gas. This resemblance suggests a similar microscopic origin based on electron-electron interactions.
\end{abstract}

\maketitle

\section{Introduction}

Electron--electron interactions and their role in electron transport are a topic of continuing interest in mesoscopic physics. 
Due to momentum conservation in electron--electron scattering processes, the latter do not influence the electron mobility unless paired with another scattering mechanism \cite{SeegerSemiconductorPhysicsIntroduction1991}. Interactions have been found to impact e.g. the conductivity of disordered systems via Friedel oscillations around screened impurities \cite{EfrosElectronelectroninteractionsdisordered1985,ZalaInteractioncorrectionsintermediate2001}, and are the key low-temperature decoherence mechanism in quantum transport experiments \cite{ImryIntroductionmesoscopicphysics2009} such as the Aharonov--Bohm effect \cite{AharonovSignificanceElectromagneticPotentials1959,WebbObservationfracheAharonovBohm1985,HansenMesoscopicdecoherenceAharonovBohm2001} or weak localization \cite{BergmannWeaklocalizationthin1984}. Recently, renewed interest in viscous effects observed in electron liquids at elevated temperatures has arisen \cite{BandurinNegativelocalresistance2016,BraemScanninggatemicroscopy2018}.

The rich variety of existing experiments includes attempts to probe electron-electron scattering by injecting non-equilibrium electrons into an equilibrium Fermi sea \cite{ZumbuhlAsymmetryNonlinearTransport2006,TaubertRelaxationhotelectrons2011}. 
Among them is a publication by Jura {\em et al.} \cite{JuraSpatiallyprobedelectronelectron2010}, which inspired the experiments to be presented in this paper. 
In this publication, the flow of electrons injected through a quantum point contact is imaged at energies above the thermal smearing of the Fermi--Dirac distribution.
Raster-scanning a locally depleting scanning gate tip above the open electron gas downstream of the injection point, the authors observed a contrast inversion of the branched electron flow signal at elevated source--drain bias voltages.
They interpreted this contrast inversion as a manifestation of electron--electron scattering in the electron gas.

Our experiments aim at finding this effect for so-called weakly-invasive tip potentials induced by the scanning gate.
In general, most scanning gate experiments (including branched electron flow measurements behind a point contact~\cite{TopinkaCoherentbranchedflow2001}) require a tip induced potential which depletes the electron gas locally (strongly-invasive regime).
However, we recently found a method to significantly enhance the sensitivity at non-depleting voltages (weakly-invasive regime) \cite{SteinacherScanninggateexperiments2018}, thus reducing the influence of the tip onto the unperturbed system.
This method utilizes a gate-defined open cavity structure~\cite{YanInterferenceEffectsTunable2017,YanIncipientsinglettripletstates2018,YanMagnetoresistanceelectroniccavity2018}, which concentrates the scattering density of states behind the quantum point contact and thereby enables scanning gate experiments at strongly reduced voltages applied to the scanning gate. 
In this paper, we operate such a structure in the nonlinear bias regime and find the interaction effects previously observed for electron injection into an open two-dimensional electron gas ~\cite{JuraSpatiallyprobedelectronelectron2010} in this modified setting.
Our finding may help to unravel the microscopic details of this effect by theoretical means beyond the explanation given in Ref.~\onlinecite{JuraSpatiallyprobedelectronelectron2010}.

\section{Sample and Experimental Setup}
\label{sec:sample_setup}

\begin{figure*}
 \includegraphics[width=\linewidth]{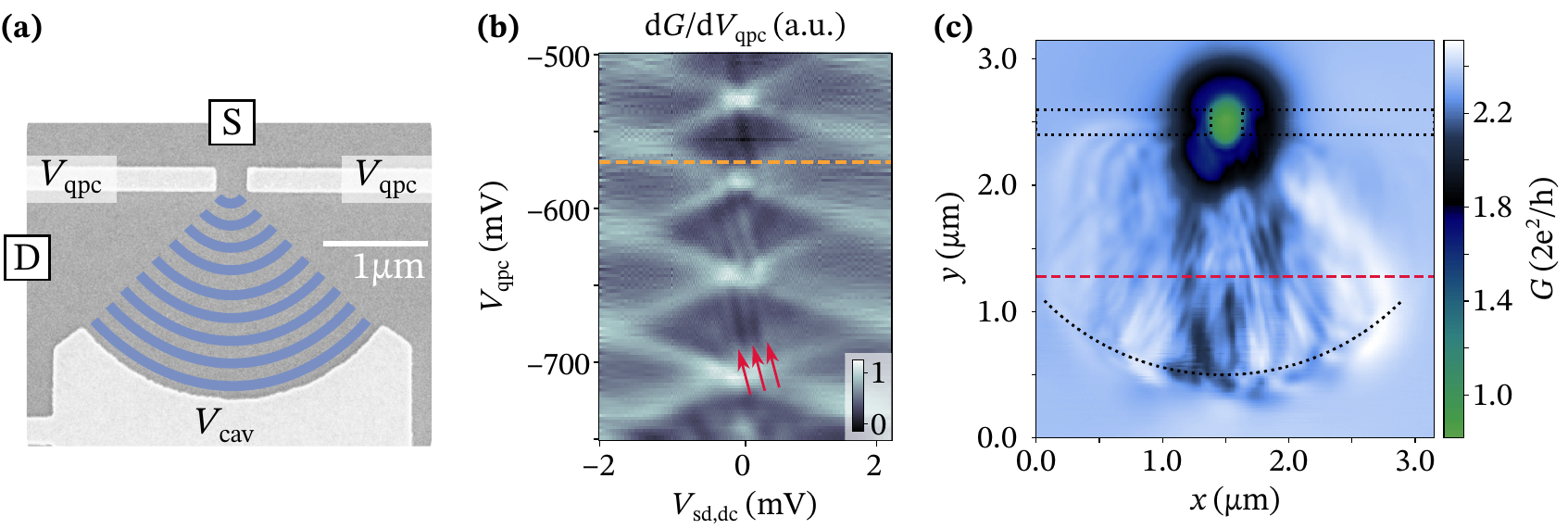}
 \caption{
 	(a) Scanning electron micrograph of the open resonator structure with Schottky gates (light gray) on a GaAs-surface (dark gray). The cavity area is depicted by the blue standing wave. The black squares denote the source (S) and drain (D) ohmic contacts. 
 	(b) Numerical derivative $dG/dV_\mathrm{qpc}$ of the differential conductance $G$ as a function of $V_\mathrm{sd,dc}$ and $V_\mathrm{qpc}$. 
 	Red arrows denote the cavity modes. 
 	The dashed orange line corresponds to the QPC-voltage used in all following measurements. The
 	data for $|V_\mathrm{sd,dc}|>\SI{1}{mV}$ is obtained with lower resolution in bias voltage and is numerically smoothened to increase the visibility of the diamond-shaped pattern.
	(c) Spatial image of the conductance $G(x,y)$ in the cavity as a function of the tip position for $V_\mathrm{tip}=\SI{-1}{V}$. Dotted lines outline the position of the Schottky gates. The data depicted in Fig. \ref{fig2: FiniteBiasTransition_withTip} is obtained along the red dashed line.}
 \label{fig1: SEM,Cav-Scan,Diamonds}
\end{figure*}

Our measurements are performed on the open resonator structure depicted in Fig.~\ref{fig1: SEM,Cav-Scan,Diamonds}(a) at temperature $T=\SI{270}{mK}$.
The sample is based on a Ga(Al)As-heterostructure [dark grey in Fig.~\ref{fig1: SEM,Cav-Scan,Diamonds}(a)]  in which a two-dimensional electron gas (2DEG) with electron density $n=\SI{1.9e11}{cm^{-2}}$ and mobility $\mu=\SI{4.38e6}{cm^2/Vs}$ is formed $\SI{90}{nm}$ below the surface.
Negative gate voltages, applied to the $\SI{300}{nm}$-wide quantum point contact (QPC) and arc-shaped cavity gate [light grey in Fig.~\ref{fig1: SEM,Cav-Scan,Diamonds}(a)], form a $\SI{2}{\mu m}$-long resonator with an opening angle of $\SI{90}{^{\circ}}$ centered around the QPC.

Applying a bias voltage $V_\mathrm{sd}=V_\mathrm{sd,ac}+V_\mathrm{sd,dc}$ between the source (S) and grounded drain (D) contact, we perform both equilibrium and non-equilibrium measurements of the differential conductance $G=I_\mathrm{sd,ac}/V_\mathrm{sd,ac}$ through the sample. Here, $I_\mathrm{SD}$ is the measured source-drain current, $V_\mathrm{sd,ac}=\SI{50}{\mu V_\mathrm{rms}}$ for all measurements, and the dc-voltage is varied between $V_\mathrm{sd,dc}=\left[\SI{-2}{mV},\SI{2}{mV}\right]$.

To explore the local properties of electron transport through the open resonator, we perform scanning gate microscopy measurements. To this end, we raster-scan a voltage-biased metallic tip approximately $\SI{65}{nm}$ above the open resonator structure while measuring the differential conductance $G(x,y)$ as a function of the tip position $(x,y)$. Unless stated otherwise, the tip is biased at a voltage $V_\mathrm{tip}=\SI{-1}{V}$, which induces a tip potential with an amplitude much smaller than the Fermi energy $E_\mathrm{F}$ \cite{SteinacherScanninggateexperiments2018}. Electrons interacting with this weakly-invasive tip-induced potential are not backscattered by a hard-wall potential~\cite{ErikssonEffectchargedscanned1996,TopinkaImagingCoherentElectron2000,TopinkaCoherentbranchedflow2001} but rather experience gentle electron lensing. 

\section{Characterization of the Cavity}

\subsection{Characterization of the cavity in absence of the SGM tip}

\label{sec: Characterization_noTip}
We first characterize the open resonator in absence of the tip by measuring the differential conductance $G(V_\mathrm{sd,dc},V_\mathrm{qpc})$.
The numerical derivative $dG/dV_\mathrm{qpc}$ of the latter is depicted in Fig.~\ref{fig1: SEM,Cav-Scan,Diamonds}(b) and exhibits the characteristic diamond-shaped pattern associated with non-equilibrium measurements of QPCs. 
The dark rhombi with $dG/dV_{\mathrm{qpc}}\approx\SI{0}{}$ correspond to regions with constant differential conductance on a conductance plateau, their extent in bias direction yielding the subband spacing $\Delta_{\mathrm{sb}}=\SI{1.5}{meV}$. 
All the following measurements are performed at $V_{\mathrm{qpc}}=\SI{-570}{mV}$ [c.f. orange dashed line in Fig. \ref{fig1: SEM,Cav-Scan,Diamonds}(b)].  
Additionally to the diamond-shaped pattern, we observe parallel and equally spaced lines in the differential conductance in the region of the QPC-plateaus [cf red arrows in Fig. \ref{fig1: SEM,Cav-Scan,Diamonds}(b)]. These lines are observed for nonzero cavity gate voltages only and are a manifestation of cavity modes with an average energy spacing $\Delta E_{\mathrm{cav}}=\SI[separate-uncertainty = true]{236(19)}{\mu eV}$.~\cite{KatinePointContactConductance1997, RosslerTransportSpectroscopySpinCoherent2015}

\subsection{Characterization of the cavity in presence of the SGM tip}
\label{sec: Characterization_tip}

To study the conductance through the cavity on a local scale, we perform SGM-measurements above the whole cavity area limited by the QPC-gates on one side and the cavity gate on the other side. 
The differential conductance $G(x,y)$ measured at different tip positions $(x,y)$ within this area is depicted in Fig~\ref{fig1: SEM,Cav-Scan,Diamonds}(c).
In agreement with previous work \cite{SteinacherScanninggateexperiments2018} it exhibits a distinct spatial structure of fine conductance modulations which emanate from the QPC radially. These conductance modulations arise due to the influence of the tip-induced potential onto the local density of scattering states, which emanate from the QPC into the cavity and are concentrated in the latter~\cite{GoldImagingsignatureslocal2020}.

The average conductance $G(x,y)$ in Fig. \ref{fig1: SEM,Cav-Scan,Diamonds}(c) is reduced with respect to the conductance of $G=3\times 2e^2/h$ on the third QPC plateau due to the capacitive action of the cavity gate and the tip on the QPC-channel.

\section{Finite bias measurements}
\label{sec: eeScattering_tip}

\subsection{Finite bias measurements in presence of the SGM tip}
\begin{figure}
	\includegraphics[width=\linewidth]{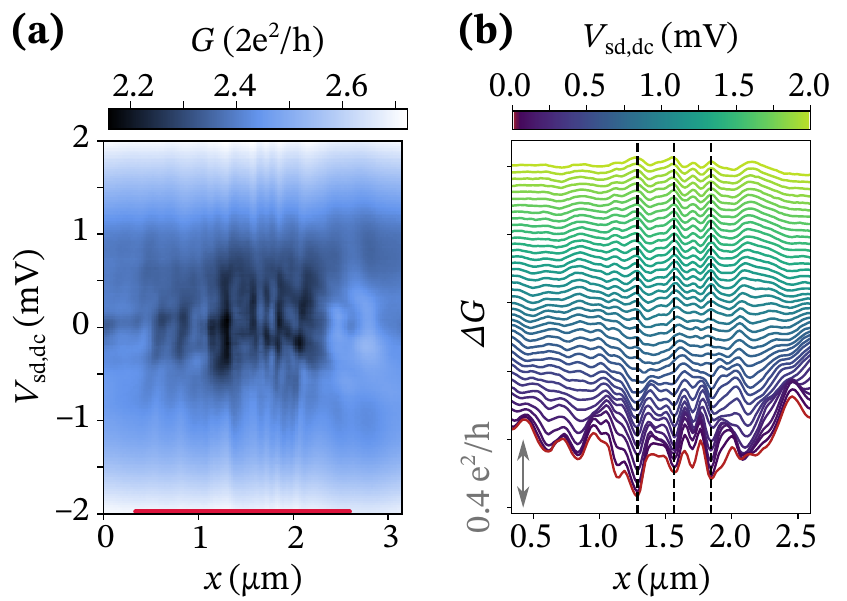}
	\caption{
		Differential conductance $G(x,V_\mathrm{sd,dc})$ along the dashed red line in Fig. \ref{fig1: SEM,Cav-Scan,Diamonds}(c). 
		(a) Raw data. 
		(b) Linewise conductance difference $\Delta G(x)$ as a function of the tip position $x$ along the red line in (a) for small steps in $V_\mathrm{sd,dc}$. The red curve denotes the conductance at $V_\mathrm{sd,dc}=\SI{0}{mV}$ and the lines are offset with respect to each other. Black dashed guides to the eye denote three exemplary minimum-to-maximum transitions.}
	\label{fig2: FiniteBiasTransition_withTip}
\end{figure}

Based on previous measurements on an open 2DEG behind a QPC ~\cite{JuraSpatiallyprobedelectronelectron2010}, we perform SGM measurements at finite source--drain voltages in order to probe e-e-scattering in the open resonator on a local scale.

The thus measured differential conductance $G(x,V_\mathrm{sd,dc}$) for tip positions $x$ along the dashed red line in Fig~\ref{fig1: SEM,Cav-Scan,Diamonds}(c) and bias voltages $V_\mathrm{sd,dc}$ is depicted in Fig.~\ref{fig2: FiniteBiasTransition_withTip}(a).

At zero source-drain-bias we recover the conductance modulations $G\in  [2.15,2.5] \cdot 2e^2/h$ already observed along the dashed red line in the spatial cavity map [cf. Fig. \ref{fig1: SEM,Cav-Scan,Diamonds}(c)]. With increasing source-drain bias $V_\mathrm{sd,dc}$, the overall conductance $G$ through the sample increases while maintaining its distinct spatial modulation. Minima (or maxima) of the conductance occur at exactly the same tip positions $x$ for source-drain biases of up to approximately $V_\mathrm{sd,dc}^\mathrm{trans.} \approx \SI{0.6}{mV}$. However, this changes significantly for bias voltages $|V_\mathrm{sd,dc}|>\SI{.6}{mV}$, at which previous maxima in the differential conductance have turned into minima and vice versa.

To emphasize this minimum-maximum transition in the differential conductance, we plot single lines ${\Delta G(x)=G(x)-\left<G(x)\right>}$ at equally spaced $V_\mathrm{sd,dc}$ between $V_\mathrm{sd,dc}=\SI{0}{mV}$ and $\SI{2}{mV}$ in Fig. \ref{fig2: FiniteBiasTransition_withTip}(b). Here, $\left<G(x)\right>$ is the conductance averaged along $x$ for fixed dc source-drain voltage $V_\mathrm{sd,dc}$. At specific fixed tip-positions $x$, the differential conductance shows a transition from minima at $|V_\mathrm{sd,dc}|<\SI{.6}{mV}$ to maxima at $|V_\mathrm{sd,dc}|>\SI{.6}{mV}$. Three of these transitions are marked by the black dashed lines in Fig.~\ref{fig2: FiniteBiasTransition_withTip}b.
A similar behavior is also observed for strongly invasive tip-potentials which induce a potential amplitude larger than the Fermi-energy in the 2DEG (see Appendix~\ref{app: Transition_StrongTip} for measurements at $V_\mathrm{tip}=\SI{-6}{V}$).

Our findings can be put into the context of available theory and the experiment in Ref.~\onlinecite{JuraSpatiallyprobedelectronelectron2010}, in which non-equilibrium carriers are injected through a QPC into an open 2DEG region. In the latter, the observed contrast inversion in regions of branched electron flow at source-drain voltages of up to $V_\mathrm{sd,dc}=\SI{2.5}{mV}$ is explained based on electron--electron (e--e) scattering.

The e-e-scattering rate $\Gamma_\mathrm{ee}$ depends on the square of the excess energy $\Delta$ above the Fermi-energy at which an electron is injected into the system ($\Gamma_\mathrm{ee}\propto \Delta^2$, see discussion below). 
Increasing the source-drain-bias $V_\mathrm{sd,dc}$ from zero to $\SI{2}{mV}$, e--e scattering in the cavity will thus become particularly relevant for electrons injected at the highest energies.

As reported in Ref.~\onlinecite{JuraSpatiallyprobedelectronelectron2010}, other inelastic scattering mechanism for hot electrons in 2DEGs (among which the most important ones are plasmon emission \cite{GiulianiLifetimequasiparticletwodimensional1982} and the excitation of acoustic phonons \cite{IhnSemiconductorNanostructuresQuantum2009}) are irrelevant at the injection energies under investigation (see Appendix A in Ref.~\onlinecite{JuraSpatiallyprobedelectronelectron2010} for details). 

\subsection{Finite bias measurements in absence of the SGM-tip}
\label{sec: eeScattering_noTip}

\begin{figure}
	\includegraphics[width=\linewidth]{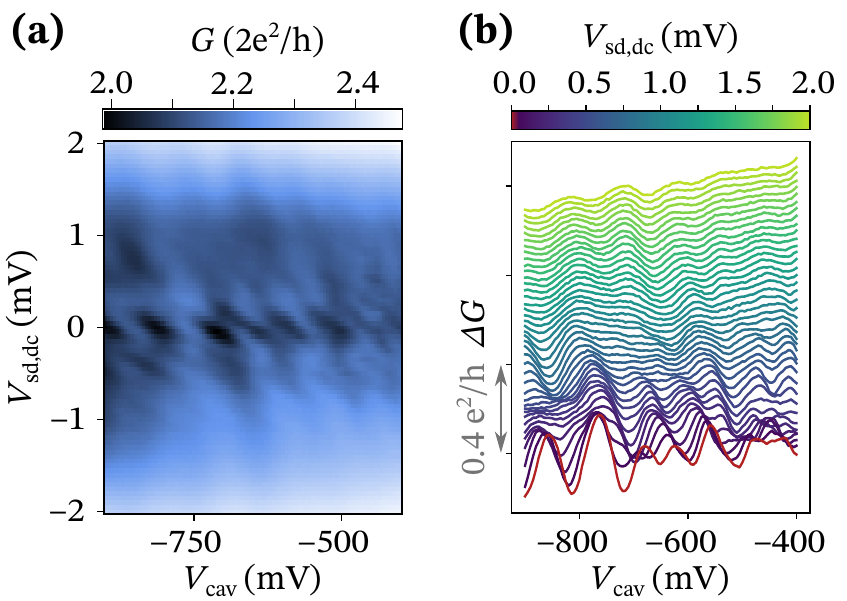}
	\caption{
		Differential conductance $G(V_\mathrm{cav},V_\mathrm{sd,dc})$ in absence of the tip. 
		(a) Raw data. 
		(b) $\Delta G(V_\mathrm{cav})=G(V_\mathrm{cav})-\left<G(V_\mathrm{cav})\right>$ for small steps in $V_\mathrm{sd,dc}$, where $\left<G(V_\mathrm{cav})\right>$ is the average conductance in $V_\mathrm{cav}$ at fixed $V_\mathrm{sd,dc}$. The conductance at zero dc-bias ($V_\mathrm{sd,dc}=\SI{0}{mV}$) is denoted in red and the lines are offset with respect to each other.}
	\label{fig3: FiniteBias_withoutTip}
	\end{figure}
	
Signatures of e--e-scattering have been shown to be present in many transport experiments in absence of a scanning tip, ranging e.g from hydrodynamic flow experiments \cite{MolenkampObservationKnudsenGurzhi1994,deJongHydrodynamicelectronflow1995,GusevViscouselectronflow2018} to Young's double slit experiments \cite{YacobyInterferencedephasingelectronelectron1991}.
Even though the tip-induced potential in our SGM-measurements is smaller than the Fermi energy, it does influence the scattering states in the cavity [cf Fig. \ref{fig1: SEM,Cav-Scan,Diamonds}(c)]. 
This raises the question of whether the observed minimum-maximum-transition is observable only in the presence of the SGM-tip or if it is an intrinsic signature of e--e-scattering in the cavity also present in the absence of the tip.

In an attempt to resolve this question, we measure the differential conductance $G(V_\mathrm{cav},V_\mathrm{sd,dc})$ in the absence of the tip for a fully formed cavity [$V_\mathrm{cav}$ below the pinch-off voltage].
The changing cavity gate voltage $V_{\mathrm{cav}}$ thereby replaces the varying tip-induced potential.
Figure~\ref{fig3: FiniteBias_withoutTip}(a) shows the resulting differential conductance. 
Again, we observe a clear transition between regions with maximal and minimal conductance but at a slightly different dc voltage $V_\mathrm{sd,dc}^\mathrm{trans,2}\approx\SI{200}{\mu V}$.
The shift of the position of the minima/maxima in $V_{\mathrm{cav}}$ arises due to the gating of the cavity modes by the bias voltage. 
This gating effect is even more obvious in Fig.~\ref{fig3: FiniteBias_withoutTip}(b) which is obtained by the same analysis that was done to obtain Fig. \ref{fig2: FiniteBiasTransition_withTip}(b) [here, the average was taken along $V_{\mathrm{cav}}$]. 
Due to this gating effect, it is impossible to identify whether minima in the conductance turn into maxima at higher bias. Figure~\ref{fig3: FiniteBias_withoutTip}(b) also shows that the shift of the minima/maxima is not linear and therefore cannot be accounted for easily. Therefore, a minimum-to-maximum-transition could not be conclusively observed in the absence of the tip.

\section{Discussion}
The electron-electron scattering length $l_\mathrm{ee}$ in the system is given as $l_\mathrm{ee}=v_\mathrm{F}\tau_\mathrm{ee}$, where $v_\mathrm{F}$ is the electron velocity at the Fermi-energy and $\tau_\mathrm{ee}$ is the electron-electron scattering time. 
The latter can be estimated to be \cite{deJongHydrodynamicelectronflow1995,GiulianiLifetimequasiparticletwodimensional1982}
\begin{equation*}
\frac{1}{\tau_\mathrm{ee}}=\frac{E_\mathrm{F}}{h}\left(\frac{\Delta}{E_\mathrm{F}}\right)^2\left[\ln \left(\frac{E_\mathrm{F}}{\Delta}\right)+\ln \left(\frac{2 q_{\mathrm{TF}}}{k_\mathrm{F}}\right) +1\right].
\end{equation*}
Here, $\Delta$ is the excess energy with respect to the Fermi-energy $E_\mathrm{F}$,
$k_\mathrm{F}$ is the Fermi wave number and $q_{\mathrm{TF}}$ is the two-dimensional Thomas-Fermi screening wave vector.
Taking the excess energy $\Delta=-|e|V_\mathrm{sd,dc}=\SI{\pm 0.6}{meV}$ at the minimum-maximum transitions [see Fig.~\ref{fig2: FiniteBiasTransition_withTip}(b)], we find an e--e-scattering length of $l_\mathrm{ee}=\SI{3.1}{\mu m}$.
This length is smaller than the path-length of the round-trip between QPC and cavity gate.
The life-time broadening of the cavity modes in the open resonator thus becomes significant with respect to the cavity mode spacing at these injection energies. This may be the reason for the decreasing amplitude of the conductance modulations in Fig.~\ref{fig2: FiniteBiasTransition_withTip}(a) with increasing $V_\mathrm{sd,dc}$.

Our measurements differ from those in Ref.~\onlinecite{JuraSpatiallyprobedelectronelectron2010} in two ways.
First, our sample consists of an open resonator formed between a quantum point contact and an arc-shaped cavity gate instead of an open 2DEG behind a quantum point contact.
Second, our measurements are obtained with tip-induced potentials lower than $E_{\mathrm{F}}$ instead of the the strongly-invasive tip potential used in the experiment by Jura {\em et al.} \cite{JuraSpatiallyprobedelectronelectron2010}.
The tip-induced potential in our experiments thus does not backscatter electrons but rather gently lenses the propagating electrons.
Surprisingly, despite these differences, our data yield a minimum-to-maximum transition at finite bias voltage, similar to the measurements in Ref.~\onlinecite{JuraSpatiallyprobedelectronelectron2010}.

Due to the complex scattering dynamics in the cavity, the exact microscopic origin of the minimum-to-maximum transition remains elusive. 
However, the qualitatively similar phenomenology of our data with the results in Ref. ~\onlinecite{JuraSpatiallyprobedelectronelectron2010}, as well as the estimates of $l_\mathrm{ee}$ given above, suggest the relevance of e--e interactions in the cavity involving the injected non-equilibrium electrons. 
Due to the design of our structure, an electron remains within the cavity for several roundtrips between the QPC and cavity-gates. 
This is consistent with the model of a collective ac motion of electrons in the cavity, which originates in the the ac electron flow injected through the QPC at energies $\Delta=- |e|V_\mathrm{sd,dc}$, as proposed in Ref.~\onlinecite{JuraSpatiallyprobedelectronelectron2010}

\section{Conclusions}
In conclusion, we measure non-equilibrium transport through an electronic cavity with scanning gate microscopy.
We observe a minimum-to-maximum transition as a function of the source-drain bias $V_\mathrm{sd,dc}$ in the differential conductance modulation caused by the tip-induced potential. 
Our measurements show that gentle electron lensing due to a tip-induced potential below the Fermi-energy \cite{SteinacherScanninggateexperiments2018} is sufficient to observe this transition. However, data taken in the absence of the tip show the relevance of the tip-induced potential for the observation of the transition.
Despite significant experimental differences, our observations are phenomenologically similar to strongly-invasive scanning gate measurements on electrons injected through a point contact into an open two-dimensional electron gas~\cite{JuraSpatiallyprobedelectronelectron2010}.
This suggests a similar microscopic origin of the minimum-to-maximum transition in both experiments, which is based on electron-electron scattering.
The detailed microscopic mechanisms of the elaborate scattering processes of electrons in the electronic cavity remains an interesting open question that will require further theoretical and experimental work.

\begin{acknowledgments}
	We thank Leonid Levitov and Vadim Khrapai for fruitful discussions and Peter M\"arki, Thomas B\"ahler as well as the staff of the ETH cleanroom facility FIRST for their technical support. We also acknowledge financial support by the ETH Zurich grant ETH-38 17-2 and the Swiss National Science Foundation via NCCR Quantum Science and Technology.	
\end{acknowledgments}

\appendix 

\section{Finite bias measurements with strongly invasive tip potentials}
\label{app: Transition_StrongTip}

\begin{figure}
	\includegraphics[width=\linewidth]{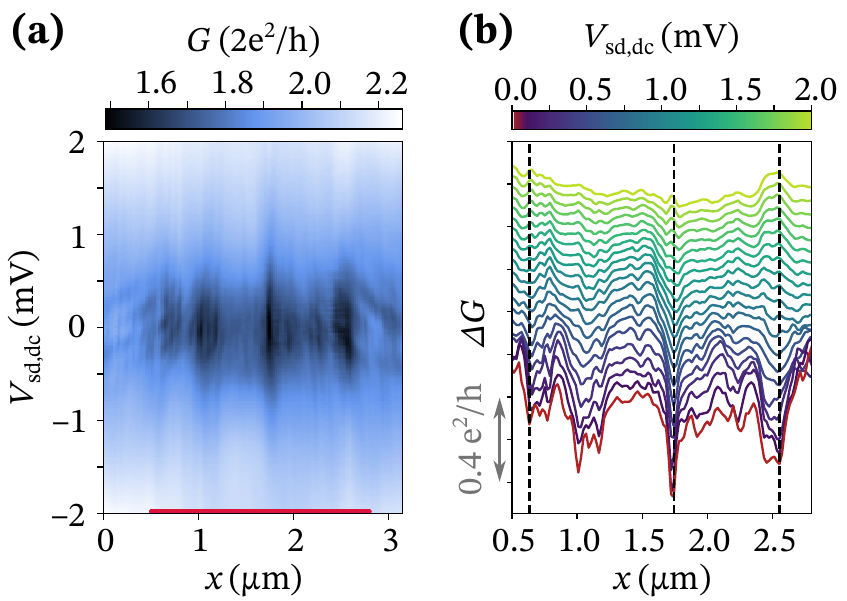}
	\caption{
		Differential conductance $G(x,V_\mathrm{sd,dc})$ for a strongly invasive tip potential ($V_\mathrm{tip}=\SI{-6}{V}$) along a similar line as the red dashed line in Fig. \ref{fig1: SEM,Cav-Scan,Diamonds}(c). 
		(a) Raw data. 
		(b) Conductance difference $\Delta G(x)$ for tip positions $x$ along the red line in (a) and small steps in $V_\mathrm{sd,dc}$. The red curve corresponds to zero source-drain bias and the lines are offset with respect to each other. Black dashed guides to the eye denote three exemplary minimum-to-maximum transitions.}
	\label{figA1: StrongTip}
\end{figure}

We evaluate the influence of a strongly invasive tip potential on the observation of the minimum-to-maximum transitions by repeating the measurement depicted in Fig.~\ref{fig2: FiniteBiasTransition_withTip} for a strongly invasive tip potential ($V_\mathrm{tip}=\SI{-6}{V}$). The thus obtained data is depicted in Fig.~\ref{figA1: StrongTip}. In accordance with previous experiments \cite{SteinacherScanninggateexperiments2018}, the additional scattering of electrons off the strongly invasive tip potential results into sharper and denser conductance modulations in the cavity area. Nonetheless, Fig.~\ref{figA1: StrongTip} depicts the same minima-maxima transition observed in Fig.~\ref{fig2: FiniteBiasTransition_withTip}. Therefore, the observation of the minima-to-maxima transitions is independent of the strength of the tip-induced potential.

\bibliography{BIB_CavFiniteBias}

\end{document}